\documentclass[twocolumn,showpacs,preprintnumbers,amsmath,amssymb,showpacs,nofootinbib]{revtex4}
\usepackage{graphicx}% Include figure files
\usepackage{dcolumn}% Align table columns on decimal point
\usepackage{bm}% bold math
\usepackage{amssymb}
\usepackage{ams}
\usepackage{amsmath}
\usepackage{color}
\usepackage{textcomp}

\def\bea{\begin{eqnarray}}
\def\ena{\end{eqnarray}}

\def\md{\mathrm{d}}
\def\tension{\frac{G\mu}{c^2}}
\def\anfac{\left|\sin\theta\right|}
\def\kepler{{\it Kepler }}
\def\corot{{\it CoRoT }}
\def\nsted{{\tt NStED }}
\def\simbad{{\tt SIMBAD}}
\begin{document}

%%%%%%%%%%%%%%%%%%%%%%%%%%%%%%%%%%%%%%%%%%%%%%%%%%%%%%%%%%%%%%%%%
\title{Observational constraints on light cosmic strings from photometry and pulsar timing}
%%%%%%%%%%%%%%%%%%%%%%%%%%%%%%%%%%%%%%%%%%%%%%%%%%%%%%%%%%%%%%%%%

%%%%%%%%%%%%%%%%%%%%%%%%%%%%%%%%%%%%%%%%%%%%%%%%%%%%%%%%%%%%%%%%%

\author{M.~S.~Pshirkov}
\affiliation{Pushchino Radio Astronomy Observatory, Astro Space Center, Lebedev Physical Institute, Pushchino, 142290, Russia\footnote[1]{e-mail: pshirkov@prao.ru}}
\author{A.V.~Tuntsov}
\affiliation{Sternberg Astronomical Institute, M.V. Lomonosov Moscow State University, 119992, Russia \footnote[2]{e-mail: tyomich@sai.msu.ru}}
%

%%%%%%%%%%%%%%%%%%%%%%%%%%%%%%%%%%%%%%%%%%%%%%%%%%%%%%%%%%%%%%%%%

\small

%%%%%%%%%%%%%%%%%%%%%%%%%%%%%%%%%%%%%%%%%%%%%%%%%%%%%%%%%%%%%%%%%
\begin{abstract}
We constrain the cosmological density of cosmic string loops using two observational signatures -- gravitational microlensing and the Kaiser-Stebbins effect. Photometry from RXTE and CoRoT space missions and pulsar timing from Parkes Pulsar Timing Array, Arecibo and Green Bank radio telescopes allow us to probe cosmic strings in a wide range of  tensions
$G\mu/c^2=10^{-16}\div10^{-10}$. We find that pulsar timing data provide the most stringent constraints on the abundance of light strings at the level $\Omega_s \sim 10^{-3}$. Future observational facilities such as the Square Kilometer Array will allow one to improve these constraints by orders of magnitude.
\end{abstract}

%%%%%%%%%%%%%%%%%%%%%%%%%%%%%%%%%%%%%%%%%%%%%%%%%%%%%%%%%%%%%%%%%

\pacs{98.80.Cq, 95.75.De, 97.60.Gb, 97.80.Jp}

\today

\maketitle

%%%%%%%%%%%%%%%%%%%%%%%%%%%%%%%%%%%%%%%%%%%%%%%%%%%%%%%%%%%%%%%%%%%%%%%%%%%%%%%%%%%%%%%%
%%%%%%%%%%%%%%%%%%%%%%%%%%%%%%%%%%%%%%%%%%%%%%%%%%%%%%%%%%%%%%%%%%%%%%%%%%%%%%%%%%%%%%%%

\section{Introduction \label{SectionI}}
Cosmic strings are now a widely recognized part of cosmological theory. Cosmic stings appear naturally in a multitude of inflationary models as topological defects from the early Universe (e.g. \cite{Allen1990,Copeland2009}, for more  studies see references in \cite{Chernoff2009}). Similar objects commonly referred to as cosmic superstrings can also be produced in fundametal string and M-theories \cite{daviskibble, polchinski}. In the present study we will not differentiate between the two classes, because their observational signatures considered in this paper are the same.

The key parameter of a cosmic string is its tension $\mu$, which is assumed to be related to the effective energy scale of
the string-producing theory $\Lambda$ by \cite{Vilenkin1994}
$$
\frac{G\mu}{c^2}\sim\frac{\Lambda^2}{M_\mathrm{Planck}^2}.
$$
Earliest theories of string formation placed time of their generation to the Grand Unification Theory epoch and therefore their tension seemed to be of order $10^{-6}$. Initially, possible tensions of string were constrained from both sides: $10^{-11}<G\mu/c^2<10^{-6}$, but eventually the lower bound was removed and  strings with arbitrarily low tension are theoretically allowed now (e.g.~\cite{Firouzjahi2005}). Cosmic strings with low tension can solve some astrophysical problems: e.g., recently, cosmic strings with tensions about $G\mu/c^2\sim10^{-12}$ were proposed as prominent source of high-energy cosmic rays \cite{Vachaspati2009}.

Simulations \citep{Allen1990, Vanchurin} suggest the energy fraction in strings has a scaling behavior: their density $\Omega_s$ (in units of critical density $\rho_0=3H_0^2/8\pi G$) does not depend on cosmological time. Recent works suggest that strings contribute a subdominant fraction to the energy balance of the Universe \cite{Sakellariadou2006}. This question has not been completely resolved yet. We will treat $\Omega_s$  essentially as a free parameter and will try to constrain it observationally.

There have been a number of attempts to limit the density of strings from various observational perspectives. Heavy enough strings, if present, would make distinctive imprints on the cosmic microwave background and shapes of distant lensed galaxies; these techniques indicate absence of strings with $G\mu/c^2\ge 10^{-7}$ in the Universe \cite{Morganson2009,Fraisse2008, sazhin}. More stringent, though more model-dependent constraints come from pulsar timing (PT): cosmic strings emit gravitational waves and the corresponding background can be detected by usual methods of PT\cite{Damour2005}. This method rules out any significant presence of cosmic strings with tensions $\mu$ at $G\mu/c^2<10^{-9}-10^{-8}$ \cite{DePies2007}. Ultimately, planned mission \textit{LISA} are expected to test the presence of gravitational wave background from lighter strings with tensions down to $G\mu/c^2\sim 10^{-14}-10^{-16}$.

String network has a complicated structure with a combination of long straight segments and a population of string loops of various lengths $L$ that were formed in the interconnections between straight strings. The loops oscillate relativistically ($\beta\sim\mathcal{O}(1)$) with an amplitude of order $L$ and period $T=L/2c$, emitting gravitational waves and eventually decaying. Only sufficiently long loops survive by the present time.

Recent simulations \cite{Chernoff2009} show that the surviving large-scale ($L\sim1\,\mathrm{pc}$ and above) loops of light cosmic strings experience considerable clustering, which closely follows that of the dark matter,  albeit with a somewhat lower amplitude. In central parts of large galaxies, such as the Milky Way, the loops' density can be enhanced by up to $10^5$ relative to its average cosmological value. This density enhancement significantly boosts the detection rates of experiments sensitive to the local population of lenses \cite{Chernoff2009, chetye}.

In this paper, we investigate how the clustering affects the prospects of detection of local cosmic strings via two observational signatures -- lensing on the string as seen in the photometry of background objects and Kaiser-Stebbins effect affecting the timing of millisecond pulsars.

The paper is organized as follows. In the next Section~\ref{depthsection}, we discuss the expected rate of the effects and lay out a simple formalism with which one can interpret the non-detection of the effect in a given experiment. Then in Sections~\ref{Eventphenomena} and~\ref{Pulsartiming} we consider lensing and Kaiser-Stebbins effect in a greater detail and calculate the constraints on the string loop population from existing observations. The final Section discusses the obtained results and offers suggestions for further research.

\section{Event rate and constraints from non-detection\label{depthsection}}

The first effect we consider is lensing of a background object by the string, in which two identical positive-parity images of the object appear on the sky when it enters a narrow strip along the string (e.g.,~\cite{Vilenkin1984, hogannarayan1984}). The width $\delta$ of the strip depends on the tension $\mu$ of the string  and the local inclination $\theta$ of the string to the line of sight:
\begin{equation}
\delta=8\pi\anfac\tension\frac{D_{\mathrm{ls}}}{D_{\mathrm{os}}} \label{stripwidth},
\end{equation}
where $D_{\mathrm{os}}$ and $D_{\mathrm{ls}}$ are the distances, respectively, from the observer to the source and from the string to the source (along the line of sight). For the presently allowed tensions of the string, the two images cannot be resolved and only an increase in the total brightness of both images is seen as the source traverses the strip.

The second observational signature of our interest is the Kaiser-Stebbins effect, which stems from the induced Doppler shift in the conic space-time around a moving string and results in a discontinuity $\Delta f$ of the observed frequency $f$ of any radiation from the source observed on the either side of the string \cite{Kaiser1984,Vachaspati1986}. It is also proportional to the string tension
\begin{equation}
\frac{\Delta\nu}\nu=8\pi\anfac\tension\beta_\perp\gamma\label{freqshift},
\end{equation}
where $\gamma=(1-\beta^2)^{-1/2}$ and $\beta_\perp$ are, respectively, the Lorentz-factor of the string and the orthogonal (to the string) component of its transverse (to the line of sight) velocity w.r.t. the source (in units of $c$). By itself, this frequency jump is again too low to be observed directly (e.g., spectroscopically) but it can manifest itself in the integrated form of pulsar timing residuals.

Both the lensing and Kaiser-Stebbins effects depend on the crossing of an observed source by a cosmic string. Thus, the probability that a given source if affected by either effect differs from the standard value of $p_0=1-\exp(-\tau_0)$ where $\tau_0$ is the optical depth given by the fraction of the sky covered by cosmic strings. However, $\tau_0$ is still useful for an estimate of the (un)importance of strips overlap in lensing.

For cosmic strings $\tau_0$ is given by summing the contributions of all infinitesimal slices of thickness~$\md D$ within a solid angle~$\md^2\Theta$ along the line of sight $D\in(0, D_\mathrm{os})$. Each contribution $\md\tau_0$ is the fraction of area covered by all strips within the volume $D^2\md^2\Theta\md D$ relative to the total area of this slice $D^2\md^2\Theta$. The area covered by all strips within the slice is given by their total length $\rho_s/\mu D^2\md D$ times the linear width $D\delta$ of strips:
\begin{equation}
\md\tau_0=\left(D^2\md^2\Theta\right)^{-1}\frac{\rho_s}\mu D^2\md^2\Theta\md D\times D\delta=\frac{\rho_s}\mu D\delta  \md D\label{ddepth}.
\end{equation}
Given that $\delta\sim\mu$ (Eq.\ref{stripwidth}), $\tau$ is actually independent of $\mu$:
\begin{equation}
\md\tau_0=8\pi\langle\anfac\rangle\frac{G\rho_s}{c^2}\frac{D D_\mathrm{ls}}{D_\mathrm{os}}\md D = 2\pi^2\frac{G\rho_s}{c^2}\frac{D_\mathrm{ol}D_\mathrm{ls}}{D_\mathrm{os}}\md D_{\mathrm {ol}} \label{deltadepth},
%\int\limits_0^{D_{os}}\rho\frac{D_{ol}D_{ls}}{D_{os}}
\end{equation}
where we used the average $\langle\anfac\rangle=\pi/4$ and added a subscript `ol' to $D$ to reflect its role in traditional lensing.
%\footnote{Since cosmic string loops are two-dimensional objects it seems natural to average $\anfac$ assuming that it is the normal to the plane of the loop rather than the direction of the string is uniformly distributed on a sphere; otherwise $\langle\anfac\rangle$ would be $1/2$, not $\pi/4$.\label{averageanfac}}

Equation~(\ref{deltadepth}) is the same as the optical depth due to a population of point lenses with mass density $2\rho\langle\anfac\rangle=\pi\rho/2$. These values have been estimated in a number of works both for the local case of MACHOs in our Galaxy and for a hypothetical cosmological population of compact lenses. As long as the density in strings does not exceed that of the dark matter, we can use the theoretical upper limits from those works to constrain $\tau_0$. The estimates ($\tau_0< 10^{-6} - 10^{-5}$  for the Galaxy and $\tau_0<10^{-2} - 10^{-1}$ in the cosmological case with source at redshift 1~\cite{eros2, ogle2, popowski, eb}) are significantly below unity. This allows us to assume that the strips do not significantly overlap in projection and at every given moment every source is affected by just one string at most.

However, the optical depth defined by~(\ref{ddepth}) is not a  measure of the expected rate of events in an observational search for cosmic strings. The latter is given by how often a given source is crossed by a cosmic string and given that by their very nature strings are extended objects, as they move across the sky they quickly sweep areas much greater than those swept by Einstein circles around point lenses with the same optical depth and velocities.

Every source swept by a string will split into two parts separated by~(\ref{stripwidth}) and their combined brightness can jump by up to a factor of 2 while the observed frequency of the source will experience a discontinuity of~(\ref{freqshift}). Whether (and how) these effects can be detected is a separate question depending on the details of a particular observational survey and discussed in the following. However, when estimating the optical depth similarly to~(\ref{ddepth}), instead of $\delta D_\mathrm{ol}$ width one should use the width of the strip swept by the cosmic string over the time span $T$ of the observational survey; this is given by $c\langle\beta_\perp\rangle T$, where $\langle\beta_\perp\rangle$ is some typical value of $\beta_\perp$ among strings contributing to the optical depth. The correct equation for the latter is then\footnote{Strictly speaking, one cannot simply add up infinitesimal contributions because nearby layers $\md D$ are not strictly independent -- that is, two neighboring layers most likely either will both have a piece of a string in them or neither will have one. However, assuming that the length of the string is small compared to the distance to the source, one can choose a `physically infinitesimal' depth of the layer $\md D$ that will make layers effectively independent.}
\begin{equation}
\label{ddepthsizesweep}
\md\tau=\frac{\rho_s}\mu c\langle\beta_\perp\rangle T\md D,
\end{equation}
which can be made very large by lowering $\mu$ due to the very poor existing lower limits on the string tension.
The probability $p$ that a change -- in either brightness of frequency of the observed source -- occurs in a given source during the observations is $p=1-\exp(-\tau)$.

The model of the experiment we consider is an observational survey searching for either a characteristic transient brightening of the source or a jump in its frequency. The time span of observations is $T$ and the number of sources monitored is $N$. The cosmic strings population is characterized by its average cosmological density $\Omega_s$ in critical units and tension $\mu$ (typical length of loops $L$ cannot be constrained with observations we consider). We assume that the string loops can cluster as suggested in~\cite{Chernoff2009} with a local enhancement $\eta$ over their average cosmological density $\Omega_s\rho_0$ so that their local density
$$
\rho_s=\Omega_s\eta\rho_0=\Omega_s\eta\frac{3H_0^2}{8\pi G},
$$
where $H_0$ is the Hubble constant; we use $\eta$ from~\cite{Chernoff2009}.

Let us now use Bayesian inference to see how a non-detection of the effect in the survey constrains the parameters of the strings $(\Omega, \mu)$. Given these parameters, we can calculate the probability that the effect happens in $i$-th source:
\begin{equation}
p_i=1-\exp\left(-\frac{\Omega_s\rho_0}\mu T c\langle\beta_\perp\rangle\int\limits_0^{D_i}\eta\md D\right)
\label{pithroughtau}
\end{equation}
integrating the enhancement $\eta$ along the line of sight.

The probability of actually detecting the effect needs to account for the limited efficiency of any survey. Even if the change in brightness or frequency does occur, we might miss it because the observations are not continuous, the effect is too weak to be observed or can hide in the intrinsic variations of the source. This is accomplished by multiplying $p_i$ by efficiency factors $\varkappa_i$ that measure the probability of detection given that the effect does happen.

Assuming that the sources are affected by strings independent of each other, the probability $Q$ of non-detection in the entire survey is then the product of the probabilities of non-detection in every source
$$
Q=\mathcal{P}\left(\mathrm{No~detection}|\Omega_s, \mu\right)=\prod\limits_{i=1}^N\left(1-\varkappa_i p_i\right),
$$
which, in the Bayesian sense, evaluates to likelihood
$$%\begin{equation}
Q(\Omega_s, \mu)=\prod\limits_{i=1}^N\left\{1-\varkappa_i\left[ 1-\exp\left(-\frac{\Omega_s\rho_0}\mu T c\langle\beta_\perp\rangle\int\limits_0^{D_i}\eta\md D\right)\right]\right\}.
%\label{likelihood}
$$%\end{equation}
%In the limit of $\tau_i\ll1$, $N\gg1$ this reduces to intuitively clear
%$$
%Q\approx\exp\left(-\sum\limits_i^N\varkappa_i\tau_i\right)=\exp\left(-\frac{\Omega_s\rho_0}\mu T %c\langle\beta_\perp\rangle\sum\limits_i^N\varkappa_i\int\limits_0^{D_i}\eta\md D\right).
%$$
The posterior distribution density of $(\Omega_s, \mu)$ set by the non-detection is then given by the Bayes theorem:
\begin{equation}
p^\prime\left(\Omega_s, \mu\right)=Q(\Omega_s, \mu)p(\Omega_s, \mu)\left[\int\md p(\Omega_s,\mu)\,Q(\Omega_s, \mu)\right]^{-1}
\label{Bayesianconstraing}
\end{equation}
where integration in the denominator extends over the parameter space measured by the prior $p(\Omega_s, \mu)$.
% which is chosen flat in $\lg\Omega$ and $\lg\mu$.

Alternatively, one can interpret $p^\prime(\Omega_s, \mu)$ as a scaled probability density of $\Omega_s$ for a fixed $\mu$ so that the probability of string density being less than $\Omega$ as a function of $\mu$ is
\begin{equation}
P(\Omega_s, \mu)=\int\limits_0^{\Omega_s}p^\prime(\Omega_s^\prime, \mu)\md\Omega_s^\prime\left[\int\limits_0^\infty p^\prime(\Omega_s^\prime, \mu)\md\Omega_s^\prime\right]^{-1}.
\label{POmega}
\end{equation}
Efficiencies $\varkappa_i$ depend not only on $(\Omega_s, \mu)$ but also on the properties of the survey, including those of individual sources. They will be calculated below for both lensing and Kaiser-Stebbins effects.

A choice of $\varkappa_i\in\{0,1\}$ is particularly convenient. Assuming that $\eta$ is constant within the probed volume of space, the likelihood function reduces to a simple form:
\begin{eqnarray}
Q(\Omega_s, \mu)=\exp\left[-\frac{\Omega_s\rho_0}\mu\eta(\mu)c\langle\beta_\perp\rangle\sum\limits_{i:\varkappa_i=1} T_iD_i\right] \label{Qcorot} \hspace{3mm} \\
=\exp\left[-\frac{\Omega_s\eta(\mu)}{G\mu/c^2}\frac{3\langle\beta_\perp\rangle H_0^2}{8\pi c}\sum\limits_{i:\varkappa_i=1}T_iD_i\right] 	 \nonumber,
\end{eqnarray}
where the sum includes only those targets, for which $\varkappa_i=1$.

For this likelihood equation~(\ref{POmega}) simplifies significantly once a flat prior in~$\Omega_s$ is chosen. In this case $Q(\Omega_s, \mu)=\exp\left[-\Omega_s/\Omega(\mu)\right]$, where the sensitivity of the experiment
\begin{equation}
\Omega(\mu)\equiv\left[\frac{\eta(\mu)}{G\mu/c^2}\frac{3\langle\beta_\perp\rangle H_0^2}{8\pi c}\sum\limits_{i:\varkappa_i=1}T_iD_i\right]^{-1}. \label{Omegamu}
\end{equation}
This value then constrains the string density $\Omega_s$ at a given confidence level $P$ according to
\begin{eqnarray}
\Omega_s(\mu, P)=-\Omega(\mu)\ln(1-P)\hspace{3.6cm}\label{Omegaconstrain}\\
=\frac{G\mu}{c^2\eta(\mu)}\frac{8\pi c\ln1/(1-P)}{3\langle\beta_\perp\rangle H_0^2}\left(\sum\limits_{i:\varkappa_i=1} T_iD_i\right)^{-1}.\nonumber
\end{eqnarray}
These equations will be used below.

\section{Lensing by Cosmic strings\label{Eventphenomena}}
\subsection{Phenomenology of string lensing}
\begin{figure}
\center
\includegraphics[width=86mm]{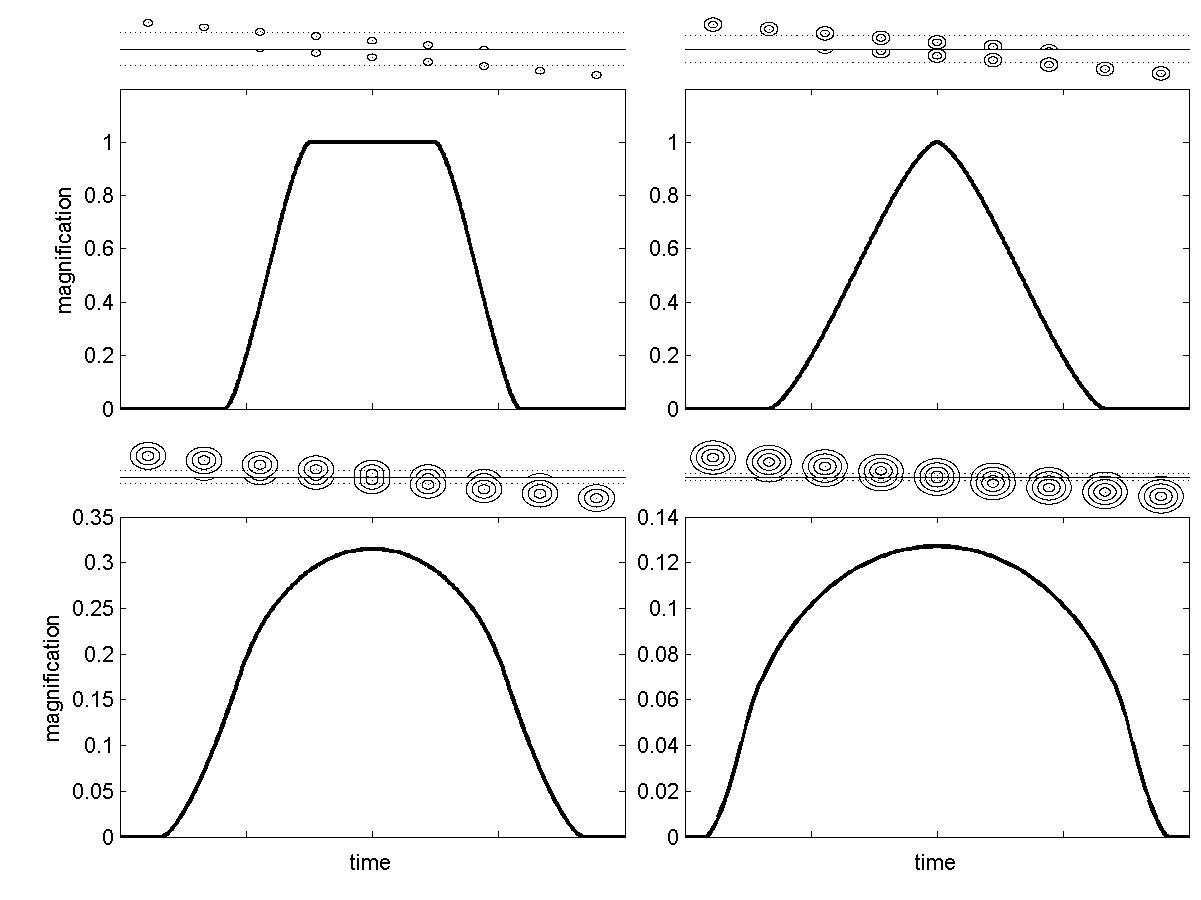}
\caption{The morphology of images and expected light curves  for cosmic strings lensing a uniform brightness disc of various sizes. From upper-left to bottom-right $r/\delta$ is $0.2, 0.5, 2, 5$. The position of the string and the strip edges are shown by solid and dotted straight lines, respectively.
}
\label{lensingpattern}
\end{figure}

Cosmic strings produce a distinct pattern of lensing (e.g.,~\cite{Vilenkin1984, hogannarayan1984}). For every point-like source entering a narrow strip of width $\delta$ along the string, a second positive-parity image appears in a `duplicate' strip on the other side of the string, the two images separated by $\delta$. As the source continues its way towards the string, the second image moves away from it until the first image disappears on contact with the string itself (see upper plots in Figure~\ref{lensingpattern}).

If the two images cannot be resolved, which is the case we consider in this paper, neither an astrometric shift in the position of the source on the sky can be detected, and what the observer sees is a temporary increase in the total brightness of the source. For a small enough source that fits into the strip completely, the brightness will increase twofold. For larger sources, the increase will be given by the total flux in the part of the source that fits within the strip.

In particular, for a disc of uniform brightness, simple geometry gives the following fractional increase in the area and flux:
\begin{equation}
f(x)=\frac1\pi\left.\left(\arcsin y+y\sqrt{1-y^2}\right)\right|_{y=\mathrm{max}(-1, -(x-\delta/2)/r)}^{y=\mathrm{min}(1, (x+\delta/2)/r)},\label{fluxincrease}
\end{equation}
where $r$ is the radius of the source and $x$ is the distance from its center to the mid-line of the strip.
%\footnote{One can clearly see that $f$ reaches its maximum as a function of $x$ when the centre of the source lies at the mid-line of the strip (cf. Fig.~\ref{lensingpattern}) -- hence the choice of origin for measuring $x$.}.

For small sources, $r\le\delta/2$ the amplitude of the effect $f_{\mathrm{max}}=1$ while for $r>\delta/2$
$$%\begin{equation}
f_{\mathrm{max}}=\frac2\pi\left[\arcsin\frac\delta{2r}+\frac\delta{2r}\sqrt{1-\left(\frac\delta{2r}\right)^2}\right]. %label{amplitudelargesources}
$$%\end{equation}
For very large sources $r\gg\delta$ the light curve and amplitude formulae %~(\ref{fluxincrease}, \ref{amplitudelargesources})
reduce to
$$%\begin{equation}
f(x)\approx\frac{2\delta}{\pi r}\sqrt{1-\left(\frac x r\right)^2} %\label{fluxincreasehugesources}
%$$%\end{equation}
~\mathrm{and}~
%$$%\begin{equation}
f_{\mathrm{max}}\approx\frac{2\delta}{\pi r}, %\label{amplitudehugesources}
$$%\end{equation}
respectively. Figure~\ref{lensingpattern} shows the morphology of lensing and the respective light curves for uniform discs of different radii.

Light curves for more complex sources would be similar and for sources smaller than the strip width the curve will be  the same as in the top left panel of Fig.~\ref{lensingpattern} except for the precise shape of the `wings' leading up to the `plateau'.
%\footnote{Note that the two `wings' need not be symmetrical for complex sources. For example, if a bright side of the source leads as it moves towards the string and the away from it, the rising wing will have a sharp rise followed by a slow climb towards the plateau while the descending wing will show a sharp drop right after the plateau followed by a slow decline towards the intrinsic brightness of the source.}.
Most importantly, the amplitude -- i.e., height of the plateau -- will be the same, $f_\mathrm{max}=1$, and this is a distinct photometrical signature for string lensing of small sources; we are not aware of any other phenomena that produce an exactly twofold increase in brightness naturally.

The time scale $t$ of such events is given by the angular velocity $\langle\beta_\perp\rangle c/D_\mathrm{ol}$ of the string with respect to the source in the plane of the sky and is given by
\begin{equation}
t=\frac{(\delta+2r_\perp)D_\mathrm{ol}}{c\langle\beta_\perp\rangle}=2\frac{D_\mathrm{ol}}{D_\mathrm{os}}\left(\pi^2\tension\frac{D_\mathrm{ls}}{c\langle\beta_\perp\rangle}+\frac{R}{c\langle\beta_\perp\rangle}\right) \label{timescale},
\end{equation}
where $2r=2R/D_\mathrm{os}$ now stands for the overall size of the source in the sky.% and $\perp$ refers to the component normal to the local direction of the string projection onto the plane of the sky.

In summary, one can see that two different regimes of lensing are possible, depending on the size of the source. For large sources $r\gg\delta$, the amplitude and time scale of the effect depend on the size of the source
\begin{equation}
f_\mathrm{max}\approx4\pi\tension\frac{D_\mathrm{ol}}{R} \hspace{5mm}\mathrm{and} \hspace{5mm} t\approx\frac{2R}{c\langle\beta_\perp\rangle}\frac{ D_{\mathrm{ol}}}{ D_\mathrm{os}}.
\label{hugesourceregime}
\end{equation}
For small sources $r\ll\delta$, the amplitude saturates while the time scale depends mostly on the string tension:
\begin{equation}
f_\mathrm{max}=1 \hspace{5mm}\mathrm{and}\hspace{5mm} t\approx2\pi^2\tension\frac{D_\mathrm{ol}D_\mathrm{ls}}{D_\mathrm{os}c\langle\beta_\perp\rangle}.
\label{smallsourceregime}
\end{equation}
For a rough estimate as to which regime applies based on the observed properties of a thermally radiating uniform source, one can utilize the Stefan-Boltzmann law to see that the angular size depends on the observed flux and temperature only: $r=T^{-2}\sqrt{F/\sigma}$. More accurate estimates need to account for the interstellar absorption and for the actual spectral energy distribution and its overlap with the detector's bandwidth. In astronomical terms, one can obtain
\begin{equation}
r\approx6.7\cdot10^{-11}\left(\frac{T_\mathrm{eff}}{10^4\,\mathrm{K}}\right)^{-2}\times10^{0.2(A-BC-(m-10))},
\label{rfromphotometry}
\end{equation}
where $m$ is the apparent instrumental magnitude of the object, $T_\mathrm{eff}$ is its effective temperature, $A$ is the interstellar absorption and $BC$ is the bolometric correction. Both $T_\mathrm{eff}$ and $BC$ can be estimated from the apparent color of the object while $A$ is well mapped in the sky.

%Let us now make a few order-of-magnitude estimates to see which regime applies in different cases of observational interest.

\subsection{Optical photometry: {\it CoRoT} and {\it Kepler}}

For hydrogen-burning, Main Sequence stars which make up the majority of the Galaxy's stellar population, the linear size $R$ varies within roughly $(10^{10} - 10^{12})\,\mathrm{cm}$ and at distances $D$ of order 1 pc to 30 kpc this corresponds to angular size $r\in(10^{-13}, 10^{-7})$; including white dwarf and red giant stars widens this interval by about an order of magnitude on both ends. Therefore, for strings with $G\mu/c^2\sim 10^{-10}-10^{-15}$ both small-source and large-source regimes may be relevant in the case of the Galaxy, though numbers tend to point towards the latter case. The amplitude of the effect could be anywhere from $10^{-7}$ to $1$ while the time scales of cosmic string lensing events within the Galaxy are rather short. Even in the extreme case, where a red giant of size $R\sim10^{13}\,\mathrm{cm}$ seen at the far end of the Galaxy ($D\sim30\,\mathrm{kpc}$) is crossed by a string of tension $G\mu/c^2\sim10^{-10}$ located half-way to the source, application of~(\ref{timescale}) yields $t$ of order an hour. For a more typical case of a main-sequence star at a few hundred pc from the Sun this reduces to tens of seconds or below.

There is some overlap of these estimates with the target sensitivity of exoplanet-hunting missions such as {\it CoRoT} and {\it Kepler}. Their target stars are relatively close to the Sun ($\sim100\,\mathrm{pc}$ for {\it Kepler} and $\sim1\,\mathrm{kpc}$ for {\it CoRoT}), which is small compared to the distance to the Galactic center. For our exploratory analysis it is therefore safe to neglect the enhancement factor $\eta$ variation within the volume probed by these stars. However, its dependence on the assumed strings tension cannot be ignored as it spans a range from $\eta\approx10^{3.6}$ at $G\mu/c^2\sim10^{-10}$ to $\eta\approx10^5 - 10^{5.2}$ at $G\mu/c^2\le10^{-13}$\cite{Chernoff2009}. Plugging these numbers into~(\ref{ddepthsizesweep}) and integrating along the line of sight, one obtains
\begin{eqnarray}
\tau=\eta(\mu)\Omega_s\frac{3H_0^2c}{8\pi G\mu}\langle\beta_\perp\rangle T D_\mathrm{os} \hspace{4.5cm} \label{depthcorot}\\
\approx 8\cdot10^{-6} \left(\frac{\eta\Omega_s}{10^{3.6}}\right)\left(\frac{\langle\beta_\perp\rangle}{0.3}\right)\left(\frac{T}{3\,\mathrm{yr}}\right)\left(\frac{D_\mathrm{os}}{10^2\,\mathrm{pc}}\right)\left(\frac{10^{-10}}{G\mu/c^2}\right) \nonumber\\
\approx 3\cdot10^{-1} \left(\frac{\eta\Omega_s}{10^5}\right)\left(\frac{\langle\beta_\perp\rangle}{0.3}\right)\left(\frac{T}{150\,\mathrm{d}}\right)\left(\frac{D_\mathrm{os}}{1\,\mathrm{kpc}}\right)\left(\frac{10^{-13}}{G\mu/c^2}\right), \nonumber
\end{eqnarray}
where the upper estimate corresponds to the survey parameters characteristic of {\it Kepler} (and high tension of strings with rather conservative consequences for $\tau$) and the lower is more relevant to {\it CoRoT} asteroseismology survey (with $G\mu/c^2$ close to the middle of the range we consider).

The number of target stars is of order $10^5$ for \kepler and $10-100$ for \corot so these estimates are not that small especially taking into account that $\Omega_s$ is poorly constrained at present. However, what is truly crucial here is the accuracy of observations given that the cosmic strings crossings are not recurring events\footnote{Strings do oscillate but the time scale of these oscillations is of order cosmic string length divided by the speed of light, which is hundred and thousands of years. Moreover, their oscillations are not restricted to any particular mode and given the plentitude of available modes there  unlikely to be any obvious periodicity in the recurrence of crossings; worse still, besides oscillations there is also center-of-mass motion of the string of unknown magnitude and direction. This makes recurrence of such events in a given source rather unlikely.} and any detection of strings would be one of those extraordinary claims that, by common wisdom, require extraordinary evidence. As a consequence, the light curve should be measured with high precision and well sampled to allow an unambiguous identification with an event described by~(\ref{fluxincrease}).

While the precision of photometry for both missions is very high, the sampling rate of \kepler is unlikely to be sufficient for our purpose. \kepler is looking at Solar-type stars with $R\sim10^{11}\,\mathrm{cm}$ and according to~(\ref{hugesourceregime}) the duration of events is $t\sim(10-20)\,\mathrm{s}$. {\it Kepler} integrates light for $30\,\mathrm{min}\sim 2\cdot10^3\,\mathrm{s}$ and therefore there is no chance to resolve the light curve of the string events, except for an unlikely case of string crossing the star almost in parallel to the string itself\footnote{Just how unlikely depends on which oscillations modes are excited on the string. It appears that $P_\beta=\beta_\perp/\langle\beta_\perp\rangle$ is a sensible upper limit to the probability that transverse velocity of the string is less than $\beta_\perp$.}. Thus, for {\it Kepler} targets $\varkappa_i\approx0$ and this mission has more or less no constraining power with regards to cosmic strings. However, when interpreting light curves of stars in search for planet transits one should keep an open eye to the possibility that an `anti-transit', in which the brightness of the star suddenly increases according to~(\ref{fluxincrease}) could be caused by a cosmic string rather than a planetary transit.

{\it CoRoT} case seems more promising\footnote{The CoRoT space mission was developed and is operated by the French space agency CNES, with participation of ESA's RSSD and Science Programmes, Austria, Belgium, Brazil, Germany, and Spain.} . In the asteroseismology mode of the satellite, the brightness of a small number of stars is sampled once a second (of which $\sim0.8\,\mathrm{s}$ is the integration time and the rest is readout etc.), although the publicly available data are integrated to $32\,\mathrm{s}$ exposures, which is optimal for  stellar seismology studies. The targets are relatively bright stars ($m_\mathrm{V}\sim 6^m - 9^m$) and despite the short exposure, their brightness is measured to a sufficiently high precision (from $\Delta f\approx0.005$ for $m_\mathrm{V}=6^m$ to $\Delta f\approx0.02$ for $m_\mathrm{V}\sim9^m$) at a reliable signal-to-noise of $S/N=10$. This is roughly the range of parameters expected for cosmic string crossings as explained in the first two paragraphs of this subsection. We therefore proceed to see how a non-detection of such events in the {\it CoRoT} data constrains the parameters of loops.

To do so, we need to specify what we would consider a reliable detection of the event -- that is, impose specific `cuts' on the parameters of events that would guarantee that it is noticed in the data and not confused with some other effect such as stellar variability. We choose the following criterion: (a) the duration of the event should be at least $t_\mathrm{min}=10\,\mathrm{s}$ so that the light curve~(\ref{fluxincrease}) is sampled at $10$ points by the \corot satellite and (b) the light curve of the crossing event should be measured with an accuracy allowing to resolve it into at least $K=5$ flux levels at a given signal-to-noise $S/N$. That is, we choose
\begin{equation}
\varkappa=\left[
\begin{array}{ll}
1, &t\ge t_\mathrm{min}\mathrm{~and~}f_\mathrm{max}/K\ge (S/N)\varepsilon_F\\
 0,&\mathrm{otherwise}
\end{array}\right. , \label{kappacorot}
\end{equation}
where $\varepsilon_F=\sigma_F/F$ is the observational relative uncertainty of the measured flux $F$, which is available from the \corot  data.
For a given source $(R, D_\mathrm{os})$ conditions in~(\ref{kappacorot}) limit the string tension from below at a value $\mu_\mathrm{min}$ that satisfies both the duration and flux increase thresholds. Assuming that the string is halfway to the source, one obtains, using~(\ref{timescale}) and~(\ref{hugesourceregime}):
%\footnote{For \corot targets, $5(S/N)\varepsilon_\mathrm{F}$ is always less than $1$, therefore $f_\mathrm{max}$ constraint from~(\ref{smallsourceregime}) is essentially irrelevant.}
\begin{equation}
\tension\ge\mathrm{max}\left(\frac{K(S/N)\varepsilon_FR}{2\pi D_\mathrm{os}}, \frac{2c\langle\beta_\perp\rangle t_\mathrm{min}-2R}{\pi^2 D_\mathrm{os}}\right) \label{mufromkappa}
\end{equation}
%A little further algebra shows that the former constraint applies when $R\ge c\langle\beta_\perp\rangle t_\mathrm{max}/[1+(\pi/4) (S/N)K\varepsilon_F]$ while the latter is more constraining for smaller sources.
The efficiencies $\varkappa_i$ are either $1$ or $0$ depending on whether~(\ref{mufromkappa}) is satisfied, thereby limiting our sensitivity at low tensions. Limitations at the other end of assumed tension range come from a gradual decrease of the enhancement factor $\eta$ for $G\mu/c^2\ge10^{-13}$. We approximate the dependence found in~\cite{Chernoff2009} with a simple broken line
\begin{equation}
\lg\eta(\mu)=\left[
\begin{array}{ll}
5.1, & \lg(G\mu/c^2)\le-13\\
5.1-0.5\left(13+\lg(G\mu/c^2)\right),&\lg(G\mu/c^2)>-13
\end{array}\right. . \label{etamu}
\end{equation}

\begin{table}
\caption{\corot target stars. The following parameters are given: the duration of the observing run~$T$, distance to the target~$D$, radius of the star~$R$, flux measurement precision~$\varepsilon_F$. Where possible, we use $D$ and $R$ quoted in either \nsted (preferably) \cite{nsteddatabase} or {\tt SIMBAD}~\cite{simbaddatabase} database; where not, we compute these quantities from the quoted effective temperature, apparent and absolute magnitudes and color~\cite{tablefootnote}. We estimated $\varepsilon_F$ as a square root of the average flux in photons because photon noise is dominant in \corot photometry of these bright stars. The last column shows the minimum tension~(\ref{mufromkappa}) probed by a given target star (assuming $t_\mathrm{min}=10\,\mathrm{s}$, $K=5$ and $S/N=10$).}
%One can see that the sensitivity of the `golden' subsample extends only down to $G\mu/c^2\ge2\cdot10^{-12}$. The minimum tension constrained by the entire sample is $4.8\cdot10^{-13}$.
\label{corottable}
\begin{tabular}{lrrrcr}
Star id & T (days) & D (pc) & R ($R_\odot$ ) & $\varepsilon_F$ (\textperthousand) & $G\mu_\mathrm{min}/c^2$\\
\hline
{\it Golden:} &&&&&\\
HD 49933 	&61	&30	&1.3	&0.32	&2.0$\cdot10^{-11}$\\
HD 51106	&61	&190	&3.4	&0.68	&2.2$\cdot10^{-12}$\\
HD 175272	&27	&85	&1.8	&0.70	&2.7$\cdot10^{-12}$\\
HD 175543	&27	&160	&2.8	&0.59	&1.9$\cdot10^{-12}$\\
HD 175726	&27	&27	&1.0	&0.48	&7.3$\cdot10^{-11}$\\
HD 181420	&157	&49	&1.6	&0.47	&2.8$\cdot10^{-12}$\\
HD 181440	&157	&160	&4.0	&0.29	&1.3$\cdot10^{-12}$\\
HD 181906	&157	&71	&1.5	&0.75	&2.9$\cdot10^{-12}$\\
%\hline
%{\it Photo-radius:} &&&&&\\
HD 50846	&58	&1100	&4.9	&1.12	&9.0$\cdot10^{-13}$\\
HD 174884	&27	&950	&2.8	&0.91	&4.8$\cdot10^{-13}$\\
HD 174966	&27	&120	&2.0	&0.81	&2.4$\cdot10^{-12}$\\
HD 175869	&27	&340	&6.9	&0.30	&1.1$\cdot10^{-12}$\\
HD 180642	&157	&480	&2.0	&1.03	&7.7$\cdot10^{-13}$\\
HD 180973	&157	&100	&3.1	&0.49	&2.7$\cdot10^{-12}$\\
HD 181231	&157	&1100	&3.1	&1.24	&6.3$\cdot10^{-13}$\\
HD 181555	&157	&110	&8.2	&0.72	&9.6$\cdot10^{-12}$\\
HD 182198	&157	&770	&4.8	&0.90	&1.0$\cdot10^{-12}$\\
HD 50170*	&61	&690	&160	&0.51	&2.1$\cdot10^{-11}$\\
HD 50747*	&61	&150	&23	&0.29	&7.9$\cdot10^{-12}$\\
HD 50890*	&55	&390	&410	&0.33	&6.2$\cdot10^{-11}$\\
HD 175679*	&27	&160	&12	&0.35	&4.8$\cdot10^{-12}$\\
HD 181907*	&157	&110	&14	&0.30	&6.8$\cdot10^{-12}$\\
%\hline
%{\it Photo-all:} &&&&&\\
HD 50773	&58	&330	&1.9	&1.82	&1.9$\cdot10^{-12}$\\
HD 50844	&58	&360	&2.7	&1.54	&2.1$\cdot10^{-12}$\\
HD 50405	&55	&330	&1.4	&1.70	&1.3$\cdot10^{-12}$\\
HD 292790	&55	&330	&3.0	&1.84	&3.0$\cdot10^{-12}$\\
HD 174936	&27	&220	&1.9	&1.23	&1.9$\cdot10^{-12}$\\
HD 175542	&27	&330	&2.1	&1.48	&1.7$\cdot10^{-12}$\\
HD 174987	&27	&330	&2.5	&1.50	&2.0$\cdot10^{-12}$\\
HD 181072	&157	&310	&2.0	&1.48	&1.7$\cdot10^{-12}$\\
%\hline
\end{tabular}
\end{table}

Table~\ref{corottable} presents $(T, D_\mathrm{os}, R, \varepsilon_F)$ for \corot target stars mined from {\tt NStED} database~\cite{nsteddatabase}. Stars are included in the table and in the analysis regardless of their known variability. We assume that intrinsic stellar variations are unlikely to present themselves because string crossings with profile like~(\ref{fluxincrease}) at these time scales; external variability is also expected to affect the flux at much greater time scales, as the velocities of potential perturbers other than strings are not relativistic.

Out 30 \corot targets, for which the data has been made public so far, only eight (top in the table) have reliable estimates of both distance and radius in the \nsted database. For the majority (14 stars, middle of the table), only an estimate of distance is available while the radius has been estimated photometrically; this is of special concern in the case of giant stars (marked with *  in the table) due to large uncertainty in estimating the bolometric correction for these stars. For eight stars (bottom of the table) both the distance and radius were estimated photometrically. We have checked numerically that giants and stars with photometrical distance estimates contribute little to the string density-constraining power of \corot data.

%\begin{figure}
%\center
%\includegraphics[width=86mm]{corotconstaint.jpg}
%\caption{Constraints on the average density of cosmic string loops from the currently available \corot data. Contours are drawn at 68\% confidence level for the complete sample of \corot targets listed in Table~\ref{corottable} and a few subsets of these stars -- the {\it Golden}, {\it Photo-radius}, {\it Photo-all} subsamples and giant stars that are marked with * in the table. The area to the left of the vertical line at $G\mu/c^2=4.8\cdot10^{-13}$ is unconstrained by these data.}
%\label{corotconstraintfig}
%\end{figure}

Figure~\ref{combinedconstraintfig} shows the constraint on $\Omega_s$ as a function of $\mu$ given by~(\ref{Omegamu}, \ref{Omegaconstrain}) assuming a non-detection of lensing signal in the light curves of stars listed in Table~\ref{corottable}. One can see that the presently available \corot data does not significantly constrain the density of cosmic string loops. This is mostly due to proximity of the source and strict cuts imposed on the lensing parameters, which severely limit the range of tension for which the lensing effect is assumed detectable. We, however, believe that these cuts are adequate given that cosmic strings have never been detected before and therefore strong evidence would be needed to lay such claims.

The envelope curve of the constraints roughly follows $\propto\mu^{-3/2}$ line with occasional wiggles due to eventual inclusion of sources with larger and large $\mu_\mathrm{min}$ into analysis. This behavior is due to the fact that the number density of strings, which controls the detection rate, is $\propto\mu^{-1}$ at a fixed mass density, which is itself $\propto\eta(\mu)\propto\mu^{-1/2}$ in the range of tensions considered. Therefore, the overall minimum of the graph is attained, {\it ceteris paribus} for those stars that have lowest $\mu_\mathrm{min}$. The latter quantity is mostly affected by the angular size of the target star and this suggests to look for effect in the smallest sources possible.

%Let us estimate the optical depth to such events as a function of $(\Omega_s, \mu)$ using source parameters similar to those of the {\it Kepler} target stars. This mission is looking after some $10^5$ Solar-type stars within $100\,\mathrm{pc}$ of the Sun. Within such scales the enhancement $\eta$ of the local density of cosmic strings relative to its average cosmological value is essentially constant. According to~\cite{Chernoff2009} $\eta$ for strings is $\sim 10^5$ for strings $G\mu/c^2\le10^{-13}$ and decreases to $\sim10^{3.5}$ for $G\mu/c^2\sim10^{-10}$.

\subsection{X-ray variability: Sco~X-1}

Some of the smallest sources one can see in the sky are high-energy astrophysics sources such as accreting neutron stars or black holes. An object of $R\sim(10-100)\,\mathrm{km}$ placed at $D\sim(1-10)\,\mathrm{kpc}$ allows one to probe strings with tensions down to $G\mu/c^2\sim10^{-16}$ still in the small-source regime. However, using accreting neutron stars and the like is complicated for two reasons. These small objects are crossed by cosmic strings in milliseconds and are often highly variable. This forces us to consider only the `ultimate' light curve with a twofold increase in brightness as candidate lensing events and focus on the brightest of sources so that this signal is reliably detectable on top of the photon noise and intrinsic variability of the source at millisecond level.

The object that best matches the above criteria is Scorpion~X-1, which is in fact the brightest persistent X-ray source in the sky not including the Sun. It is a low-mass X-ray binary at $D\approx2.8\,\mathrm{kpc}$ away, in which a neutron star accretes matter from a low-mass companion. The size of the X-ray emitting region of the source is believed to be just that of the neutron star and can be safely assumed to be $2R\le100\,\mathrm{km}$, which corresponds to $r\sim10^{-15}$ in the sky. Observations of this source by the Proportional Counter Array (PCA) onboard {\it Rossi X-ray Timing Explorer} ({\it RXTE}) satellite with a total exposure time $T=3.32\cdot10^{5}\,\mathrm{s}$ were used by~\cite{scox1nat} to discover trans-neptunian objects in the Solar system by occultations they produce in the light curve of Sco~X-1. The duration of such events is of order milliseconds, which happens to be very suitable for our purposes.

The flux of Sco~X-1 observed by {\it RXTE} PCA is $F\approx10^5\,\mathrm{counts}/\mathrm{s}$, which allows one to detect doubling of flux at a signal-to-noise level $S/N=10$ on time scales $t=(S/N)^2/F=10^{-3}\,\mathrm{s}$. The intrinsic variability is known to be Poissonian for Sco~X-1 and at these time scales is significantly below {\it RXTE} instrumental fluctuations (see \cite{vanderklis} and figure~1 thereof in particular). The time scale $t=10^{-3}\,\mathrm{s}$ corresponds to the string crossing time for a source of $2R\approx100\,\mathrm{km}$ and is less than that for all strings with tension $G\mu/c^2\ge G\mu_\mathrm{min}/c^2=10^{-16}$, which are also those for which doubling of the flux occurs. Therefore, we expect that a twofold increase in the flux of Sco~X-1 would be reliably detected had it happened during the observations.

However, no such events were observed in the light curve (cf. figure~1 of~\cite{scox1nat}).
%\footnote{In fact, the flux was even more stable than within a factor of two and one could in principle probe even lighter strings with these data; however, we conservatively consider only the `ultimate' light curves.}
Application of~(\ref{Omegamu}) then yields the following constraints on the density of cosmic string based on the {\it RXTE} data for Sco~X-1:
\begin{equation}
\Omega(\mu)=5.6\cdot 10^{-2}\left(\frac{G\mu/c^2}{10^{-16}}\right)\left(\frac{10^5}{\eta(\mu)}\right).
\label{ScoX1Omegamu}
\end{equation}
The constraints from {\it RXTE} observations of Sco~X-1 are plotted in figure~\ref{combinedconstraintfig}. One can see that this data provides competitive upper limits on the density of loops of the lightest cosmic strings.

\section{Search with Pulsar timing\label{Pulsartiming}}
Usually pulsar timing (PT) is used to  constrain the presence of strings in rather indirect way: oscillations of string loops produce specific gravitational wave background and its influence can be sought in the anomalous residuals of millisecond pulsars. In this paper a different effect is considered. We look for anomalous residuals caused by the string crossing of the line of sight to the pulsar via the Kaiser-Stebbins effect.

Such crossing instantaneously changes the apparent frequency of the pulsar by a small amount $\Delta\nu$ according to~(\ref{freqshift}). Observationally, this effect is the same as that of the gravitational wave burst with memory except that the latter applies to all observed pulsars whereas the former is specific to the pulsar, whose line of sight was crossed by the string. Observational signatures of gravitational wave bursts with memory were thoroughly studied in \cite{Pshirkov2009} and we will extensively utilize that analysis in the present work.

\begin{figure}
\center
\includegraphics[width=86mm]{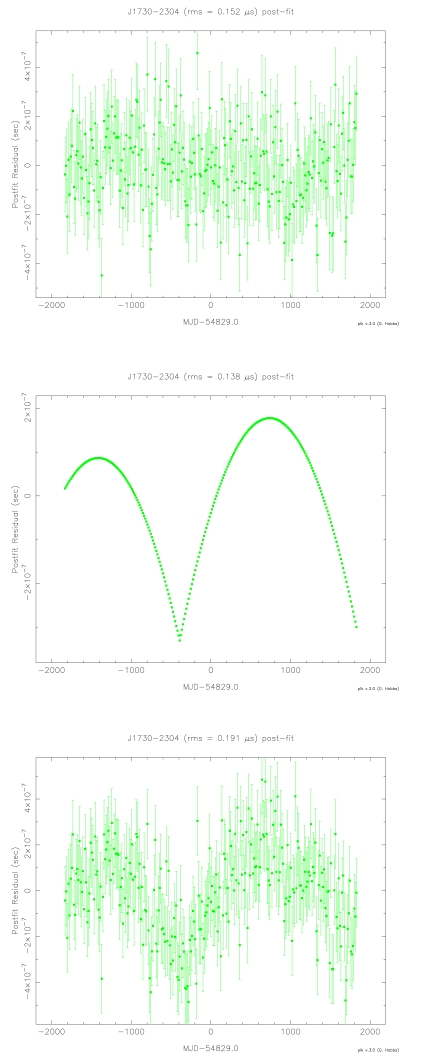}
\caption{ Pulsar timing residuals generated with the \textsc{fake} plugin of \textsc{tempo2} package \cite{Hobbs2006}. The first picture shows 10 year  span of pulsar timing with $\sigma= 150~\rm ns$, the second one present impact of string with $G\mu/c^2=10^{-15}$ crossing at the fourth year of observations (timing noise omitted); the third one shows this impact on `real' pulsar with $\sigma= 150~\rm ns$}
\label{residualsfig}
\end{figure}

The shift in frequency manifests itself in the pre-fit pulsar timing residuals as a broken line function that is equal to zero before the moment of crossing and grows linearly after that. Standard reduction procedures of pulsar timing data necessarily includes fitting of residuals for a priory unknown frequency and its first derivative, which absorb  linear and quadratic trends correspondingly  in post-fit residuals (see the middle panel of fig.~\ref{residualsfig}).

We will assume that the string crossing is detected if the amplitude $\delta s$ of the residual due to the Kaiser-Stebbins effect exceeds twice the root mean square (rms) of the pulsar timing noise $\sigma$:
\bea \delta s \sim 2 \sigma,\label{level_of_detection}.\ena
Using eqs. (18-20) of \cite{Pshirkov2009}, we arrive at:
\bea\delta s \approx\frac{3}{64}\frac{\Delta \nu}{\nu}T \label{Kappapulsar_0}, \ena
where $T$ is the total time span of observations of the pulsar.

This assumption corresponds to the following choice of $\varkappa$:
\begin{equation}
\varkappa=\left[
\begin{array}{ll}
1, &s\ge 2\sigma\\
 0,&\mathrm{otherwise}
\end{array}\right. , \label{kappapulsar}
\end{equation}
which is equivalent (cf. \ref{Kappapulsar_0}) to
\begin{equation}
\varkappa=\left[
\begin{array}{ll}
1, & G\mu/c^2\ge 64\sigma/(3\pi^2T)\\
 0,&\mathrm{otherwise}
\end{array}\right. . \label{kappapulsar_2}
\end{equation}

\begin{table}
\caption{Pulsar timing array target pulsars including PPTA and results from Arecibo ($^*$)  and Arecibo+GBT($^{**}$). The following parameters are given: the rms of measured pulsar timing noise $\sigma$, distance to the target~$D_{\mathrm {os}}$ \cite{atnfpulsars}, total time span of observations~$T$; we also list tension thresholds~$\mu_\mathrm{min}$ probed by the pulsar as given by~(\ref{kappapulsar_2}).}
\label{pulsartable}
\begin{tabular}{lcccc}
Pulsar name& $\sigma$ ($\mu s$) & $D$ (kpc) & $T$ (yr) & $G\mu_\mathrm{min}/c^2\cdot10^{15}$\\
\hline

J0437-4715&	0.2&	0.16&4.3 &	3.3\\
J0613-0200& 1.1&	0.48&5.5 &	14\\
J0711-6830&	1.6&	1.04&4.4 &	26\\
J1022+1001&	2.2&	0.40&5.5 &	28\\
J1024-0719&	1.3&	0.53&5.5 &	17\\
J1045-4509&	3.0&	3.24&5.2 &	41\\
J1600-3053&	1.0&	2.67&5.5 &	13\\
J1603-7202&	1.9&	1.64&5.5 &	24\\
J1643-1224&	1.7&	4.86&5.4 &	22\\
J1713+0747&	0.5&	1.12&5.5 &	6.0\\
J1730-2304&	1.9&	0.51&4.6 &	29\\
J1732-5049&	3.5&	1.81&5.5 &	45\\
J1744-1134&	0.8&	0.48&5.5 &	10\\
J1824-2452&	1.7&	4.90&3.1 &	39\\
J1857+0943$^*$&	1.0&	0.91&8&	12\\
J1909-3744&	0.6&	1.14&5.5&	8\\
J1939+2134$^{**}$&	$\sim$2.0&	17&8.33&8\\
J2124-3358&	2.4&	0.25&3.8 &45\\
J2129-5721&	1.2&	2.55&5.5 &	15\\
J2145-0750&	1.1&	0.50&4.3 &	18\\
\end{tabular}
\end{table}

We are now ready to compute the constraints on the cosmic string loop density $\Omega_s$ according to~(\ref{Omegamu}, \ref{Omegaconstrain}) based on the pulsar timing data collected in the Parkes Pulsar Timing Array (PPTA)  \cite{Hobbs2009} project as well as from Arecibo \cite{Kaspi1994} and Green Bank Telescope (GBT) \cite{Lommen2002} observations. The details of these data are listed in Table~\ref{pulsartable}. The resulting constraints are shown in figure~\ref{combinedconstraintfig}, which shows that current PT observations limit the density of light cosmic strings quite significantly.

\section{Results and Discussion}

\begin{figure}
\center
\includegraphics[width=86mm]{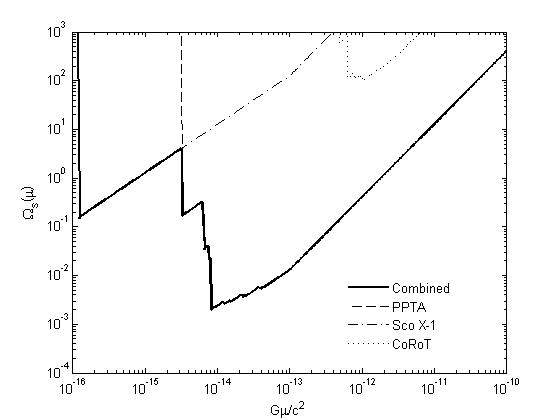}
\caption{Combined constraints on the average density of cosmic string loops based on \corot, {\it RXTE} and {\it PPTA} data at P=95\% level. One can see that pulsar timing currently provides the strongest constraints down to $G\mu/c^2\sim 10^{-14}$ while lighter strings are constrained by the available data on Sco~X-1; \corot data do not significantly constrain the population of loops at present.}
\label{combinedconstraintfig}
\end{figure}

Figure~\ref{combinedconstraintfig} presents our final result including constrains from pulsar timing, X-ray data for Sco~X-1 and precision photometry from \corot. One can see that the most stringent limits come from pulsar timing except for the lightest strings where competitive constraints are provided by {\it RXTE} data. Presently available \corot data do not have much constraining power in the range of tensions we consider. Existing observations allow one to limit the average density of cosmic string loops down to $\Omega_s\sim10^{-3}$ at $G\mu/c^2=10^{-14}$. For larger tensions the limits become eventually weaker, proportionally to $\mu^{-1}$ for $G\mu/c^2 \le 10^{-13}$ and
$\mu^{-1.5}$  for $G\mu/c^2 > 10^{-13}$ because of lesser enhancement of density  of heavy strings.

These results clearly demonstrate that the density enhancement in the Galaxy improves the limits on cosmic string abundance that can be set by test sensitive to the local population of loops as suggested by~\cite{Chernoff2009}. However, the same approach can also be applied to extragalactic sources, most notably quasars. The probability of lensing by string loops for these sources  is lower due to absence of enhancement in the extragalactic case. On the other hand, the huge distance to the source  can easily compensate for the relative deficit of string density. Moreover, these observations are insensitive to theoretical uncertainty in predicting $\eta$.

As an example we consider the 'Einstein cross' quasar QSO~2237+0305. The Optical Gravitational Lensing Experiment ({\it OGLE}) data for this source currently span more than 7 years of high-quality photometry sampled every few days~\cite{ogle2237} and no obvious signature of string lensing are seen in these light curves\footnote{The light curve for image~B does show an approximately twofold increase in brightness starting near $JD\sim2453500$ but present data is far from being sufficient to draw any firm conclusions. Unfortunately, no observations can resolve this source at present to the level needed to rule out the possibility of lensing by cosmic string.}. This non-detection can be used to place some limits on the density of cosmic strings in the range of tensions $3\cdot10^{-12}<G\mu/c^2<3\cdot10^{-10}$ (corresponding to the strings with crossing time equal to the sampling interval and data span, respectively).

Using (\ref{Omegaconstrain}) with enhancement  factor $\eta=1$, $D\sim2\,\mathrm{Gpc}$, $T\sim7.5\,\rm years$ yields
\bea
\label{huchra_2}
\Omega_s\sim4\cdot10^{-2}\left(\frac{G\mu/c^2}{3\cdot10^{-13}}\right)
\ena
These limits are inferior to constraints from pulsar timing by about an order of magnitude.

Future observational projects will noticeably improve the limits established in this paper. In \cite{Kuijken2008} it was proposed to use upcoming large-scale observational surveys of quasar variability to search for cosmic strings. In line with this paper, we might assume that a large number $N\sim10^3$ of quasars is monitored for $T\sim10\,\mathrm{years}$ on a daily basis. Non-detection of string crossings in such survey would limit the average cosmic string loop density with $G\mu/c^2\ge 10^{-13})$
at the level of
\bea
\label{qso_survey_constraint}
\Omega_s= 2\cdot10^{-5}\left(\frac{G\mu/c^2}{10^{-13}}\right)
\left(\frac{1\,\mathrm{Gpc}}{D}\right)\left(\frac{10^3}{N}\right)\left(\frac{10~\rm yr}{T}\right),
\ena
which is orders of magnitudes better that constraints presented in this paper --  essentially as a by-product of AGN long-time variability study.

Other avenues to better constraining $\Omega$ at lower tensions exist. With the eventual arrival of the Square Kilometer Array (\textit{SKA}) we can expect further enhancement  in the sensitivity due to the increase of number of pulsars and timing of more distant pulsars~\cite{ska}. Pulsar timing array at \textit{SKA} will consist of $\sim$100 pulsars, that would be timed with precision better than 100 ns. Exact limits will depend on distances  of pulsars, but we can forecast that accessible range and density estimates will improve by orders of magnitude.

%%%%%%%%%%%%%%%%%%%%%%%%%%%%%%%%%%%%%%%%%%%%%%%%%%%%%%%%%%%%%%%%%%%%%%%%%%%%%%%%%%%%%%%%
%%%%%%%%%%%%%%%%%%%%%%%%%%%%%%%%%%%%%%%%%%%%%%%%%%%%%%%%%%%%%%%%%%%%%%%%%%%%%%%%%%%%%%%%

%%%%%%%%%%%%%%%%%%%%%%%%%%%%%%%%%%%%%%%%%%%%%%%%%%%%%%%%%%%%%%%%%%%%%%%%%%%%%%%%

\section*{Acknowledgements}
The authors thank  M.V.~Sazhin and K.A.~Postnov for useful discussions and fruitful suggestions. We also thank J. P. W.~Verbiest for useful advices about \textsc{fake} plugin of \textsc{tempo2} package.  The work of M. P. is supported by RFBR Grant  No. 07-02-01034a. This research has made use of data obtained through the High Energy Astrophysics Science Archive Research Center Online Service, provided by the NASA/Goddard Space Flight Center, photometry from the ESA's \corot space mission. We thank OGLE team for making their data publicly available. Use of data from \simbad and \nsted databases  and NASA's Astrophysics Data System is gratefully acknowledged.

%%%%%%%%%%%%%%%%%%%%%%%%%%%%%%%%%%%%%%%%%%%%%%%%%%%%%%%%%%%%%%%%%%%%%%%%%%%%%%%%%%%%%%%%
%%%%%%%%%%%%%%%%%%%%%%%%%%%%%%%%%%%%%%%%%%%%%%%%%%%%%%%%%%%%%%%%%%%%%%%%%%%%%%%%%%%%%%%%

%%%%%%%%%%%%%%%%%%%%%%%%%%%%%%%%%%%%%%%%%%%%%%%%%%%%%%%%%%%%%%%%%%%%%%%%%%%%%%%%%%%%%%%%
%%%%%%%%%%%%%%%%%%%%%%%%%%%%%%%%%%%%%%%%%%%%%%%%%%%%%%%%%%%%%%%%%%%%%%%%%%%%%%%%%%%%%%%%


\begin{thebibliography}{99}
\bibitem{Allen1990}
B. Allen and E. P. S. Shellard, Phys.~Rev.~Lett., \textbf{64}, 119 (1990).
\bibitem{Copeland2009}
E.J.~Copeland and T.W.B.~Kibble, arXiv:0911.1345.
\bibitem{Chernoff2009}
D. F. Chernoff, arXiv:0908.4077.
\bibitem{daviskibble}
A.-C.~Davis and T.W.B.~Kibble, Contemp.~Phys., \textbf{46}, 313 (2005).
\bibitem{polchinski}
J. Polchinski, Int.~J.~Mod.~Phys.~A {\bf 20}, 3413 (2005).
\bibitem{Vilenkin1994}
 A.~Vilenkin and E.P.S.~Shellard, \textit{Cosmic strings and other topological defects}, Cambridge University Press (1994).
\bibitem{Firouzjahi2005}
H.~Firouzjahi, S.H.~Tye, J.~Cosm.~Astrop.~Phys.,  \textbf{03}, 009 (2005).
\bibitem{Vachaspati2009}
T.~ Vachaspati, arXiv:0911.2655 (2009).
\bibitem{Vanchurin}
V.~Vanchurin, K.D.~Olum, and A.~Vilenkin, Phys.~Rev.~D, \textbf{72}, 063514 (2005).\\
V.~Vanchurin, K.D.~Olum, and A.~Vilenkin, Phys.~Rev.~D, \textbf{74}, 063527 (2006).
\bibitem{Sakellariadou2006}
M. Sakellariadou, Ann.~Phys., \textbf{15}, 264 (2006).
\bibitem{Morganson2009}
E.~Morganson, P.~Marshall, T.~Treu, T.~Schrabback, and R.D.~Blandford,  arXiv:0908.0602.
\bibitem{Fraisse2008}
A.A. Fraisse, C. Ringeval, D.N. Spergel, and  F.R. Bouchet, Phys.~Rev.~D, \textbf{78}, 043535  (2008).
\bibitem{sazhin}
M.V.~Sazhin, O.S.~Khovanskaya, M.~Capaccioli, G.~Longo, M.~Paolillo, G.~Covone, N.A.~Grogin, and E.J.~Schreier, Mon.~Not.~R.~Astron.~Soc., \textbf{376}, 1731 (2007).\\
M.~Sazhin \textit{et al.}, Mon.~Not.~R.~Astron.~Soc., \textbf{343}, 353 (2003).
\bibitem{Damour2005}
T.~Damour and A.~Vilenkin, Phys.~Rev.~D, \textbf{71}, 063510 (2005).
\bibitem{DePies2007}
M.R.~Depies and  C.J.~Hogan, Phys.~Rev.~D, \textbf{75}, 125006 (2007).
\bibitem{chetye}
D.F.~Chernoff, S.-H.H.~Tye, arXiv:0709.1139.
\bibitem{Vilenkin1984}
A.~Vilenkin, Astrophys.~J.~Lett., \textbf{282}, 51 (1984) .
\bibitem{hogannarayan1984}
C.~Hogan and R.~Narayan, Mon.~Not.~R.~Astron.~Soc., \textbf{211}, 575 (1984).
\bibitem{Kaiser1984}
N.~Kaiser and  A.~Stebbins, Nature, \textbf{310}, 391 (1984).
\bibitem{Vachaspati1986}
T. Vachaspati, Nucl.~Phys.~B, \textbf{277}, 593 (1986).
\bibitem{eros2}
C.~Hamadache \textit{et al.}, Astron.~Astrophys., \textbf{454}, 185 (2006).
\bibitem{ogle2}
T.~Sumi \textit{et al.}, Astrophys.~J., \textbf{636}, 240 (2006).
\bibitem{popowski}
P.~Popowski \textit{et al.}, Astrophys.~J., \textbf{631}, 879 (2005).
\bibitem{eb}
N.W.~Evans and V.~Belokurov, arXiv:astro-ph/0411222.
\bibitem{nsteddatabase}
http://nsted.ipac.caltech.edu/
\bibitem{simbaddatabase}
http://simbad.u-strasbg.fr/
\bibitem{scox1nat}
H.-K.~Chang, S.-K.~King, J.-Sh.~Liang, P.-Sh.~Wu, L.~Ch.-Ch.~Lin, and J.-L.~Chiu, Nature, \textbf{442}, 660 (2006).
\bibitem{vanderklis}
M.~van~der~Klis, R.A.D.~Wijnands, K.~Horne, and W. Chen, Astrophys.~J.~Lett., \textbf{481}, 97 (1997).
\bibitem{Pshirkov2009}
M.S.~Pshirkov, D.~Baskaran, and K.A.~Postnov, arXiv:0909.0742 (2009).
\bibitem{Hobbs2006}
G.B.~Hobbs, R.T.~Edwards, and  R.N.~Manchester, Mon.~Not.~R.~Astron.~Soc., \textbf{369}, 655 (2006)
\bibitem{Hobbs2009}
G.B.~Hobbs \textit{et al.}, Publ.~Astron.~Soc.~Aust., \textbf{26}, 103 (2009).
\bibitem{atnfpulsars}
R. N.~Manchester, G.B.~Hobbs, A.~Teoh, and M.~Hobbs, Astron.~J., \textbf{129}, 1993 (2005).\\ http://www.atnf.csiro.au/research/pulsar/psrcat
\bibitem{Kaspi1994}
V.M.~Kaspi, J.H.~Taylor, and M.F.~Ryba, Astrophys.~J., \textbf{28}, 713 (1994).
\bibitem{Lommen2002}
A.N.~Lommen, arXiv/astro-ph:0208572.
\bibitem{ogle2237}
A.~Udalski \textit{et al.}, Acta~Astron., \textbf{56}, 293 (2006).
\bibitem{Kuijken2008}
K.~Kuijken, X.~Siemens, and T.~Vachaspati, Mon.~Not.~R.~Astron.~Soc., \textbf{384}, 161 (2008).
\bibitem{ska}
R.~Smits, M.~Kramer, B.~Stappers, D.R.~Lorimer, J.~Cordes, and A.~Faulkner, Astron.~Astrophys., \textbf{493}, 1161 (2009).
\bibitem{flower}
P.J.~Flower, Astrophys.J., \textbf{469}, 355 (1996).
\bibitem{tablefootnote}
The radius in Solar units is calculated from the absolute magnitude $M_\mathrm{V}$ in V and effective temperature $T_\mathrm{eff}$ according to $$ \frac{R}{R_\odot}=10^{-0.2(M_\mathrm{V}-M_\mathrm{V}^\odot) - 0.2(BC-BC_\odot)}\left(\frac{T^\odot_\mathrm{eff}}{T_\mathrm{eff}}\right)^2. $$ We use the bolometric corrections from~\cite{flower}. We compute distance $$ D=10\,\mathrm{pc}\times10^{0.2(m_\mathrm{V}-M_\mathrm{V}-A_\mathrm{V})}. $$ In two cases, for stars HD~181231 and HD 182198, Table~\ref{corottable} quotes calculated values despite the presence of an estimate of $D$ in \nsted database. The reason for this is that the distance quoted by \nsted has a very large uncertainty and significantly exceeds photometric estimate thereby making the latter conservative.

\end{thebibliography}
\end{document}